# N-P-N Bipolar Action in Junctionless Nanowire TFET: Physical Operation of a Modified Current Mechanism for Low Power Applications

Morteza Rahimian[a] ,Morteza Fathipour[1,b]

[a,b] School of Electrical and Computer Engineering, University of Tehran, P. O. Box 14395-515, Tehran, Iran

[1]Corresponding author: Email: mfathi@ut.ac.ir

*Abstract*- **In this paper we study the device physics of a technique for realizing an n-p-n bipolar transistor action in the source side of a junctionless nanowire tunneling FET (BJN-TFET). In the on-state, tunneling of electrons from valence band of the source to conduction band of the channel enhances the hole concentration as well as the potential in the source region which drives a built-in BJT transistor by forward biasing the base-emitter junction, with the source acting as a p-type region. Owing to the sharp switching of the JN-TFET and high BJT current gain, the overall performance is improved, including favorable high on-state current ($2.17 \times 10^{-6}$ A/µm), and sub 60 mV/dec subthreshold swing (~ 50 mV/dec) at low supply voltages. This approach modifies the current mechanism owning to the triggered BJT and makes the proposed structure more attractive for scaling requirements in future low power application.**

*Key Words*- Junctionless nanowire TFET (JN-TFET); Bipolar junction transistor (BJT), band to band tunneling (BTBT); on-state current; subthreshold swing; 2D TCAD simulation.

# 1. Introduction

Junctionless nanowire transistors (JNTs) are promising candidates for low power applications with incessant down-scaling in CMOS integrated circuits **[1-13]**. JNTs have genuine advantages in fabrication processing such as easiness of source and drain formation without any need for thermal budget. They are uniform and homogenous doped transistors without any metallurgical junction due to absence of sharp doping gradients which are expected to facilitate fabrication process and be free from problems associated with random dopant fluctuation **[1-13]**. However, JNTs suffer from a thermal limit of 60 mV/decade on the subthreshold swing (*SS*) and need a high supply voltage for achieving *SS* < 60 mV/dec **[6-10]**.

So, in recent years, there has been an increasing desire to explore novel devices that can provide low subthreshold swing at low $V_{DD}$ **[14-19]** while maintaining a low $I_{off}$ **[20-22]**. This dictates the need to break the 60 mV/dec barrier in the JNTs which is an obstacle for future scaling of the supply voltage **[6-10]**. Nevertheless, the major bottleneck of the JNTs that fundamentally degrades the *SS* value is drift-diffusion mechanism.

Thus, with the goal of replacing the JNTs by devices based on a new carrier injection mechanism different from diffusion over a potential barrier, the tunneling field effect transistors (TFETs) have been proposed to tackle the limitation of the JNTs and be immune to subthreshold degradation at short channel lengths **[14-19]**. However, certain drawbacks of the TFETs impede further high speed applications and adversely affect the functionality of circuits based on these transistors including: **1)** it has been demonstrated that TFETs alone suffer from an unacceptably low $I_{on}$ and cannot bring the $I_{on}$ to that of the MOSFET levels **[23-26]**. However, several methods have been attempted to surmount this concern **[27-38]. 2)** TFET also comes along with its unique property of conduction for both high negative and high positive gate voltages which restrict its utility in digital circuit design **[37-42]**. Electric field reduction, depletion region width extension, and use of large bandgap heterostructure, all on the drain side have been proposed as innovative remedies for suppression of ambipolar conductivity **[37-42]**. 3) For efficient tunneling in a TFET, an abrupt high doped junctions using complex high thermal processes are required which are not easy to access due to the dopant atoms diffusion **[43, 44]**. Hence, attention is shifting towards substitutional solution for junction limitation.

Recently, Ghosh *et al* **[45-47]** introduced a device architecture called JN-TFET which exhibited tremendous potential as it combined the advantages of TFET with sub 60 mV/dec subthreshold swing and JN-FET device **[45-48]**. This concept, although alleviates problems with random dopant diffusion, leaves problems with both low $I_{on}$ and ambipolar conduction intact.

In this paper, we propose a novel device which combines the benefits of simple fabrication process of a JN-TFET with high $I_{on}$ of a tunneling triggered bipolar junction transistor (BJT) (BJN-TFET) and we will study how the BJT is responsible for the activation of the device on-state. We show that, in the on-state of the BJN-TFET structure, the tunneling of electrons from $E_V$ of the source region to $E_C$ of the channel leads to generating holes and leaving them behind in the source region. The accumulation of holes raises the potential of the $p^+$ source region and turns on the BJT device by forward biasing the base-emitter junction with the $p^+$ source acting as a base, resulting in a high on-state current. So, abrupt transition between the on- and off-states and sub-60 mV/dec subthreshold swing are achieved in the BJN-TFET structure compared to the conventional JN-TFET (CJN-TFET) and conventional TFET (C-TFET) structures at low operating voltages. Although, bipolar-enhanced tunneling has been studied previously as a leakage-causing effect in conventional JNTs **[49]**, here it provides a mechanism for a sharp-switching device with high $I_{on}$ and sub-60 mV/dec subthreshold swing which paves the way for realizing high performance JN-TFET required for low power and low cost applications.

## 2. Device Parameters and Simulation Modeling

In this numerical study for the BJN-TFET, CJN-TFET and C-TFET structures, we have assumed the following parameters: silicon film thickness ($t_{si}$) = 10 nm, gate dielectric (SiO$_2$) thickness ($t_{ox}$) = 1 nm, main gate length ($L_{MG}$) = 30 nm.

**Fig. 1** shows schematic cross section view of the conventional JN-FET (CJN-FET) with uniform and homogenous n$^+$ doped ($N_D$ = 1×10$^{19}$ cm$^{-3}$), CJN-TFET, BJN-TFET, and C-TFET structures. In order to induce holes in the n$^+$ doped region and convert it into a p$^+$ source region, a metallic contact with an appropriate workfunction ($WF_{PS}$ = 5.93 eV) is utilized which is called here as PS-Gate (PS). Converting a part of the n$^+$ doped source into a p$^+$ source region, leaves an n$^+$ doped region between the source contact and p$^+$ source region as shown in **Fig. 1(c)**. This provides an n-p-n BJT device which is responsible for the activation of the device on-state. Worth noting that the main gate workfunction is 4.6 eV.

The 2D numerical device characteristics are carried out using Atlas from Silvaco **[50]**. Worth noting, the local BTBT model calculates a recombination-generation rate at each point solely on the field value local to that point. Thus if a sufficiently high electric field exists within a device, local band bending may be sufficient to allow electrons to tunnel from the valence band into the conduction band. To model the tunneling process more accurately, we need to take into account the spatial variation of the energy band **[33-47]**. In contrast to local tunneling models, the nonlocal BTBT model describes a physical picture of carrier transport through the barrier. This model considers the energy band profile along the entire tunneling path and thus assumes that the tunneling takes place on a series of 1D slices through the junction. However, the nonlocal BTBT model is less suitable for lightly doped p-n junctions as well as for application to the unipolar, high-field regions of a device. Thus we need to use local and nonlocal BTBT model simultaneously.

To take into account the recombination effects, we consider the Shockley-Read-Hall (SRH) model. Also due to the presence of interface traps at the silicon/oxide interface and high carrier concentration, we employ direct recombination model (Auger) **[51].** Band gap narrowing model (BGN) is used due to the high doping density in all regions of the device. The simulation mobility models take into account both field dependent as well as concentration dependent mobility.

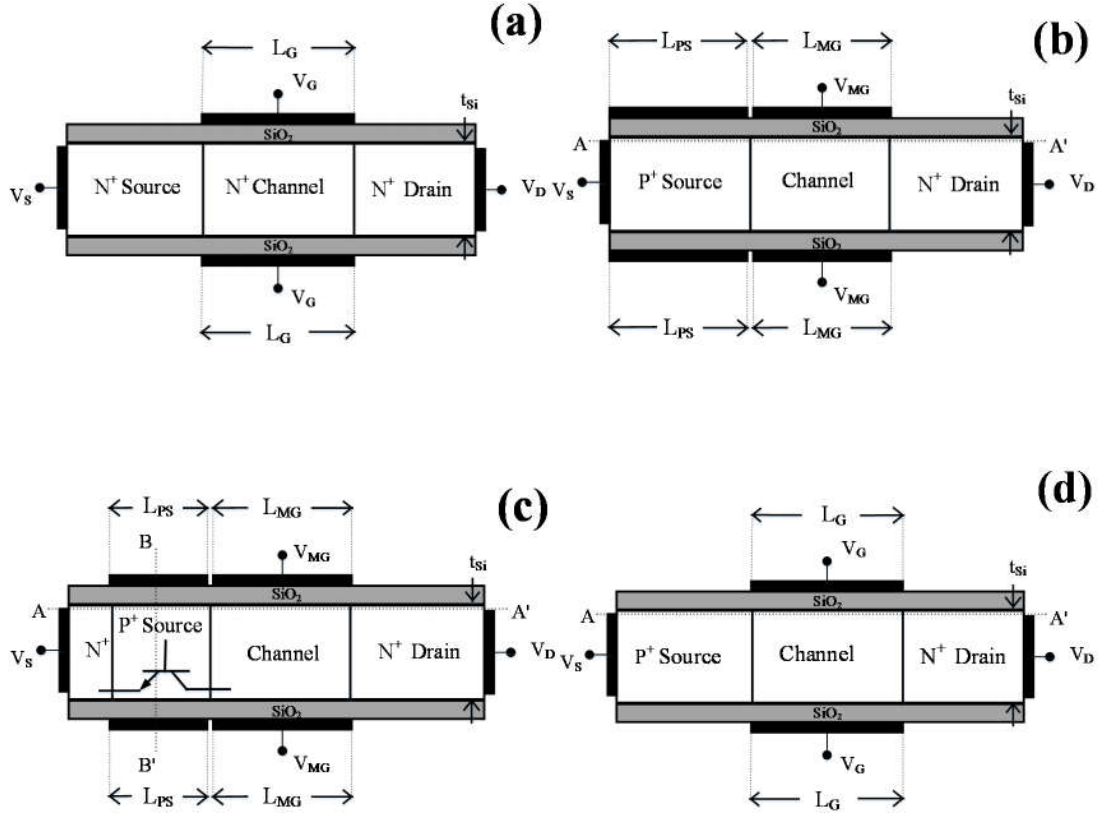

**Fig. 1**. Cross sectional view of the (a) C-JNT, (b) CJN-TFET, (c) BJN-TFET, (d) and C-TFET structures. The workfunction for the Main gate and PS-Gate are 4.6 eV and 5.93 eV, respectively. An n-p-n bipolar junction transistor action is responsible for the activation of the device on-state in the BJN-TFET structure.

### 3. Physical Operation of the BJN-TFET

In this section we describe the device physics of the triggered bipolar action in the BJN-TFET structure, provide optimized values for parameters of the proposed structure and discuss their influence on device `performance. Also we explore the short-channel effect and their impact into the sub-20 nm regimes.

### 3.1) Device physics of the BJN-TFET

**Fig. 2** demonstrates the lateral band diagram of the both structures shown in **Fig. 1(a)** in the off-state ($V_{GS}$= 0 V, $V_{DS}$= 1.0 V) and on-state ($V_{GS}$= 1.0 V, $V_{DS}$= 1.0 V) along the AA' cutline located at 0.1 nm below the Si-SiO$_2$ interface. Our goal in this paper is to realize the barrier in the source region to merge diffusive and BTBT mechanisms for carrier transport which means the BJN-TFET structure cannot purely be a tunneling FET. The band profiles along the source-channel regions of the BJN-TFET structure are similar to the bands in an n-p-n BJT with a region under the PS-Gate acting as a base and a barrier exists between the n$^+$ and p$^+$ in the source region. Such a configuration is known to degrade the off-state of the CJN-FET **[49]**.

In the off-state, the tunneling width at the collector-base junction is large which suppresses the tunneling current, leading to a negligible emitter-collector current. However, in the on-state, the tunneling width at the collector-base junction becomes small. This initiates tunneling of electrons from valence band of the p$^+$ region (base of the built-in BJT) to the conduction band of the channel (collector of the built-in BJT), leaving holes behind in the p$^+$ source region. This forward biases the base-emitter junction and eventually turns-on the n-p-n BJT, resulting in a large drain current (collector current of the BJT).

**Fig. 3** demonstrates the electron and hole concentration of the BJN-TFET structure in the off-state ($V_{GS}$= 0 V, $V_{DS}$= 1.0 V) and on-state ($V_{GS}$= 1.0 V, $V_{DS}$= 1.0 V) along the AA' cutline. As can be seen from the **Fig. 3 (b)**, in the on-state, thanks to the positive gate voltage, the electron concentration of the channel is enhanced. This provides an n-p-n pattern of a BJT device in the source-channel regions.

The transfer characteristics ($I_{DS}$-$V_{GS}$) of the BJN-TFET, CJN-TFET and C-TFET structures are compared in **Fig. 4(a)**. As it is clear from the figure, the BJN-TFET not only exhibits a significant enhancement in the $I_{on}$, but also its *SS* value is improved noticeably over a wide range of drain current compared to the CJN-TFET and C-TFET structures as is shown in **Fig. 4(b)**. Also, in **Fig. 4 (a)**, reduction of threshold voltage ($V_{TH}$) in the BJN-TFET demonstrates the responsibility of BJT action for the earlier turn-on of the device.

Worth noting, for the range of $I_{DS}$ from $2.0 \times 10^{-10}$ A/μm to $5.0 \times 10^{-9}$ A/μm, the *SS* of the BJN-TFET, CJN-TFET and C-TFET varies from 50 mV/dec to 58 mV/dec, 68 mV/dec to 100 mV/dec, and 69 mV/dec to 106 mV/dec, respectively. This exhibits the capability of the BJN-TFET structure for providing *SS*< 60 mV/dec. However, as the BJN-TFET is not operated solely by tunneling mechanism, the *SS* < 60 mV/dec is observed only for very low gate voltage, where the drain current is extremely small. This is the point at which the parasitic BJT begins to be triggered and the threshold is linked to the activation of the BJT on-state which leads to an increase in drain current. Notably, at sufficiently high gate voltages, the drift-diffusion current is much bigger than the tunneling current due to the BJT action and thereby the major component of drain current is diffusive. Thus we cannot expect the *SS* to be sub-60mV/decade.

As can be seen from **Fig. 4 (a)**, not only the $I_{on}$ current but also the $I_{off}$ current of the BJN-TFET is larger than that of the CJN-TFET structure. However the amount of $I_{on}$ enhancement is larger than the $I_{off}$ degradation which leads to higher $I_{on}/I_{off}$ ratio in the BJN-TFET structure compared to the CJN-TFET. Thus the BJN-TFET presents higher $I_{on}$ and $I_{on}/I_{off}$ ratio compared to the CJN-TFET which is desirable.

Also, there are several reported techniques that not only enhance $I_{on}$, but also increase $I_{off}$ such as employing Ge or SiGe in the source and channel regions **[26-27, 38-40, 54]**, using high-k gate dielectric **[30-36]**, n$^+$ source pocket **[15, 28, 52-53]**, strained silicon film **[55-56]**, dual material gate **[57-58]**, and some innovative fabrication methods **[16-18, 31]**. These approaches are utilized to address the $I_{on}$ concern while they suffer from $I_{off}$ degradation.

On the other hand, there are several solutions for $I_{off}$ reduction **[32, 37-42, 48, 57-58]**. As we have known, in TFETs, large tunneling width at the drain side leads to negligible tunneling probability which enables low $I_{off}$ to be achieved [**32, 37-42, 57-58**]. If we merge the techniques for $I_{on}$ increment with those for $I_{off}$ reduction, we will have a device with high $I_{on}$, low $I_{off}$ and high $I_{on}/I_{off}$ ratio which is favorable for low power applications [**32, 37-41, 57-58**].

In this paper, in order to extend the tunneling width at the drain side and also alleviate $I_{off}$ current, the drain doping is reduced. The impact of drain doping on the $I_{DS}$-$V_{GS}$ characteristic is illustrated in **Fig. 4 (c)**. As can be seen from the figure, lower drain doping results in $I_{off}$ suppression while the effect of drain doping on the $I_{on}$ is almost negligible.

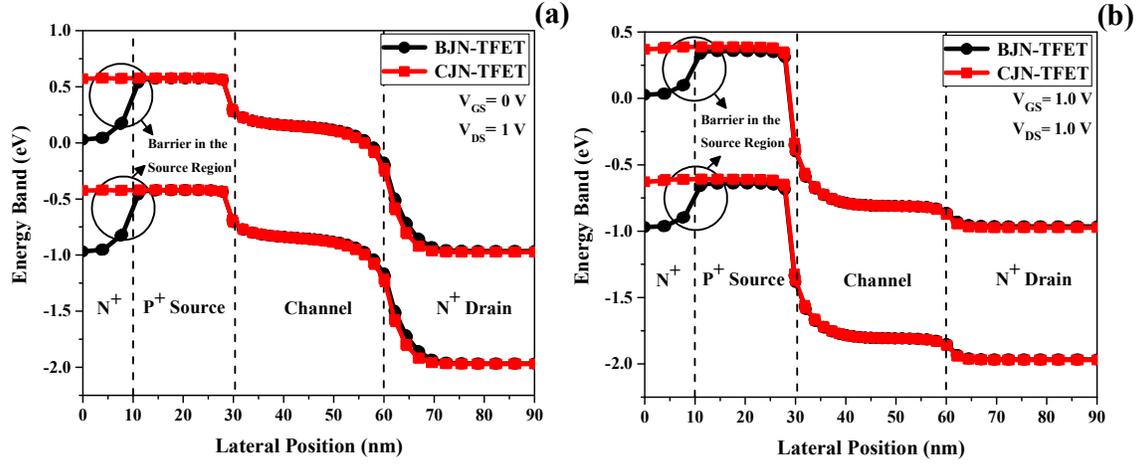

**Fig. 2.** Lateral band diagram of the BJN-TFET structure in the (a) off-state ($V_{GS}$= 0.0 V, $V_{DS}$= 1.0 V) and (b) on-state ($V_{GS}$= 1.0 V, $V_{DS}$= 1.0 V). There is a barrier between the n$^+$ and p$^+$ in the source region as well as the band diagram along the channel-source regions is similar to an n-p-n BJT transistor.

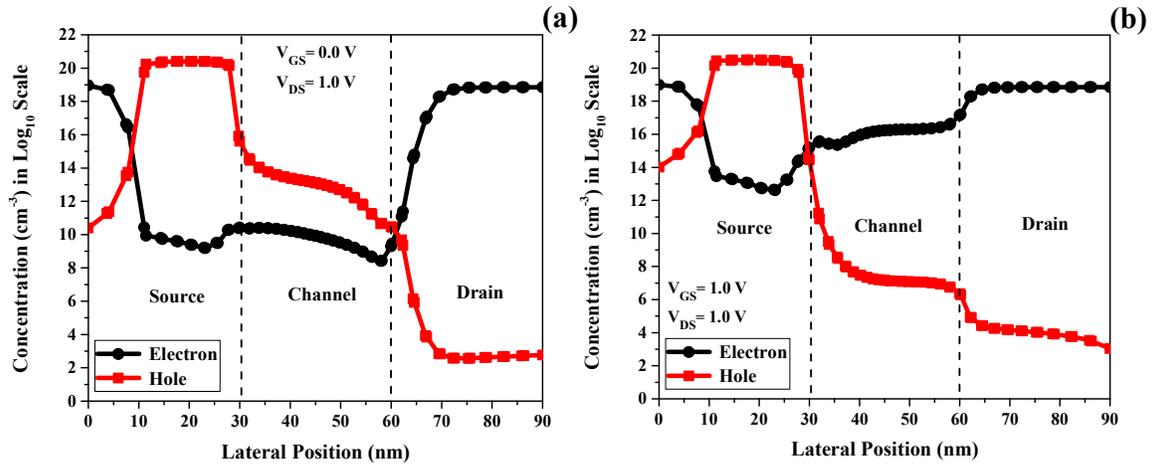

**Fig. 3.** Electron and hole concentration of the BJN-TFET along lateral direction in the (a) off-state ($V_{GS}$= 0.0 V, $V_{DS}$= 1.0 V) and (b) on-state ($V_{GS}$= 1.0 V, $V_{DS}$= 1.0 V). There is an increase in the electron concentration of the channel region when positive gate voltage is applied which provide an n-p-n pattern of a BJT device.

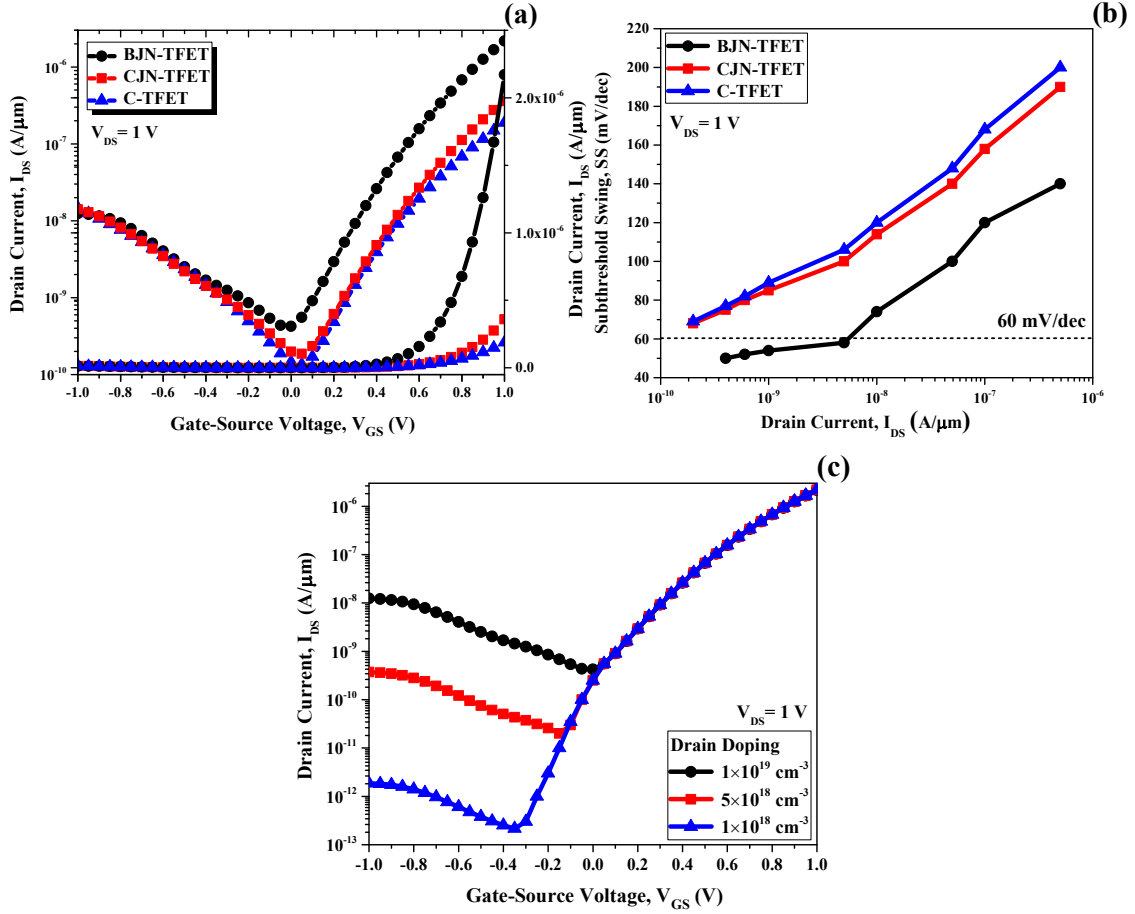

**Fig. 4.** Comparison of (a) transfer characteristic ($I_{DS}$-$V_{GS}$) and (b) $SS$-$I_{DS}$ for the BJN-TFET, CJN-TFET and C-TFET structures at $V_{DS}$= 1.0 V. The $I_{on}$, $I_{off}$, and $I_{on}/I_{off}$ of the BJN-TFET are 2.17×10$^{-6}$ A/µm, 4.24×10$^{-10}$ A/µm and 5.11×10$^3$ whereas for the CJN-TFET structure are 3.59×10$^{-7}$ A/µm, 1.99×10$^{-10}$ A/µm and 1.80×10$^3$, respectively at $V_{DS}$= 1.0 V. Threshold voltage is reduced in the BJN-TFET structure. The subthreshold swing of the BJN-TFET is improved significantly over wide range of the drain current compared the CJN-TFET and C-TFET structures. (c) Transfer characteristic ($I_{DS}$-$V_{GS}$) of the BJN-TFET structure with various drain doping. $I_{off}$ alleviation is achieved due to the lower drain doping.

**Fig. 5** compares the output characteristics ($I_{DS}$–$V_{DS}$) of the BJN-TFET, CJN-TFET and C-TFET structures when $V_{GS}$ varies from 0 V to 2 V. As can be seen from the figure, the $I_{DS}$ in the BJN-TFET structure is larger than that achieved in the CJN-TFET and C-TFET structures.

Notably, we observe that the structures exhibit clear exponential and saturation regions of operation. The saturation region in the output characteristics is because of the progressively less dependence of tunneling width on $V_{DS}$ as the drain voltage is increased.

Also, huge variation in current as gate voltage increases, shows high gate control. When low gate voltage is applied, the tunneling barrier is still high enough and hence the probability of tunneling of charge carriers would be very low. On applying higher gate voltages, continuous improvement in $I_{on}$ is observed which is due to the continuous reduction in the tunneling barrier width that increases the probability of tunneling of more and more charge carriers.

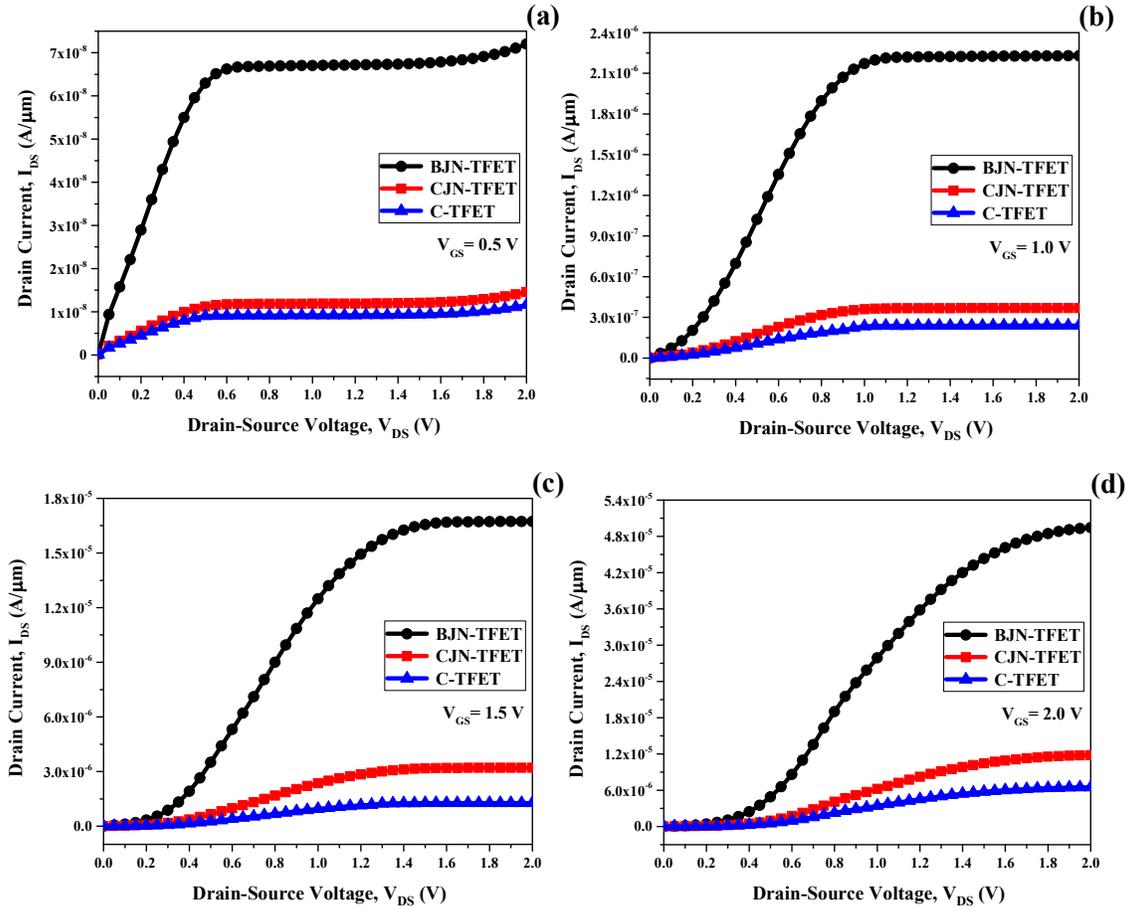

**Fig. 5.** Comparison of output characteristic ($I_{DS}$-$V_{DS}$) of the BJN-TFET, CJN-TFET and C-TFET structures at (a) $V_{GS}$= 0.5 V (b) $V_{GS}$= 1.0 V (c) $V_{GS}$= 1.5 V (d) $V_{GS}$= 2.0 V. The $I_{DS}$ of the BJN-TFET structure is larger than that achieved in the CJN-TFET and C-TFET structures. The structures exhibit clear exponential and saturation regions of operation.

To study the underlying physics of the BJN-TFET structure in detail, we have conducted device simulation both with and without the inclusion of the BTBT model in **Fig. 6(a)**. We observed that the BJN-TFET device does not turn on without considering the BTBT model which exhibits that the BTBT is responsible for the activation of the device. Also in **Fig. 6(b),** a comparison between the lateral band diagram in the on-state ($V_{GS}$= 1 V, $V_{DS}$ = 1 V) demonstrates that there is a noticeable reduction in the $n^+/p^+$ barrier in the source region when the BTBT model is applied which activates the diffusive injection. As we called earlier, tunneling of electrons from the source region into the channel leads to holes accumulation. This increases the potential under the PS-Gate and thereby reduces the barrier between the $n^+/p^+$ regions in the source region. Thus, the physics of transport in the BJN-TFET structure is an interplay of both diffusive and BTBT mechanisms.

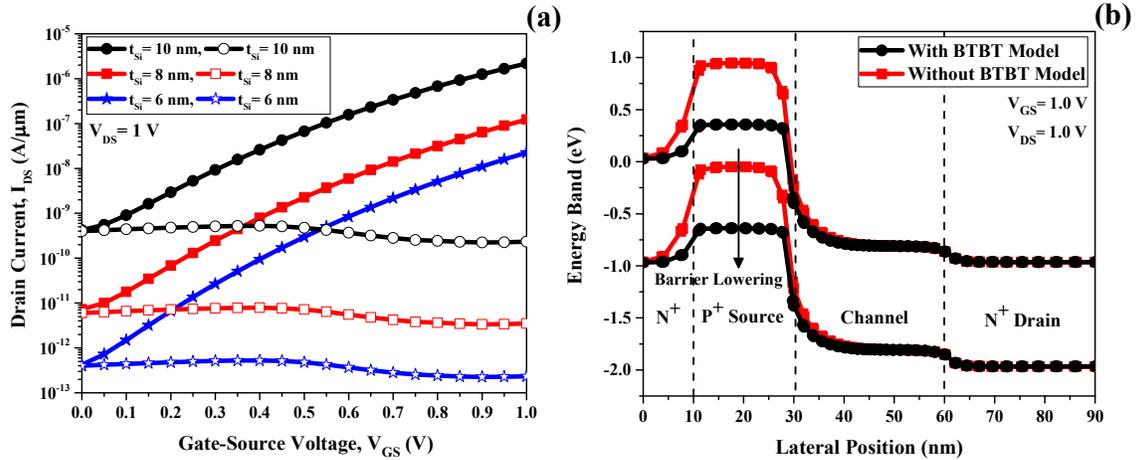

**Fig. 6.** (a) Transfer characteristics ($I_{DS}$-$V_{GS}$) of the BJN-TFET at $V_{DS}$ = 1.0 V simulated with and without BTBT model. The BTBT is responsible for turning on the BJN-TFET device. (b) Lateral band diagram of the BJN-TFET at $V_{GS}$ = 1.0 V and $V_{DS}$ = 1.0 V simulated with and without BTBT model. With inclusion of the BTBT model, there is a significant reduction in the $n^+/p^+$ barrier in the source region.

In **Fig. 7**, to further investigate this finding, we have plotted hole concentration both with and without the BTBT model in the on-state ($V_{GS}$= 1.0 V, $V_{DS}$= 1.0 V). Notably, in **Fig. 7 (c)** and **(d)**, we exhibit the hole concentration in the lateral and vertical directions along the AA' and BB' cut lines, respectively. The BB' cutline is located at 18 nm from left side of the structure. We observe a large difference in the hole concentration under the PS-Gate owing to the electrons tunneling from the source region to the channel when the BTBT model is included in the simulation.

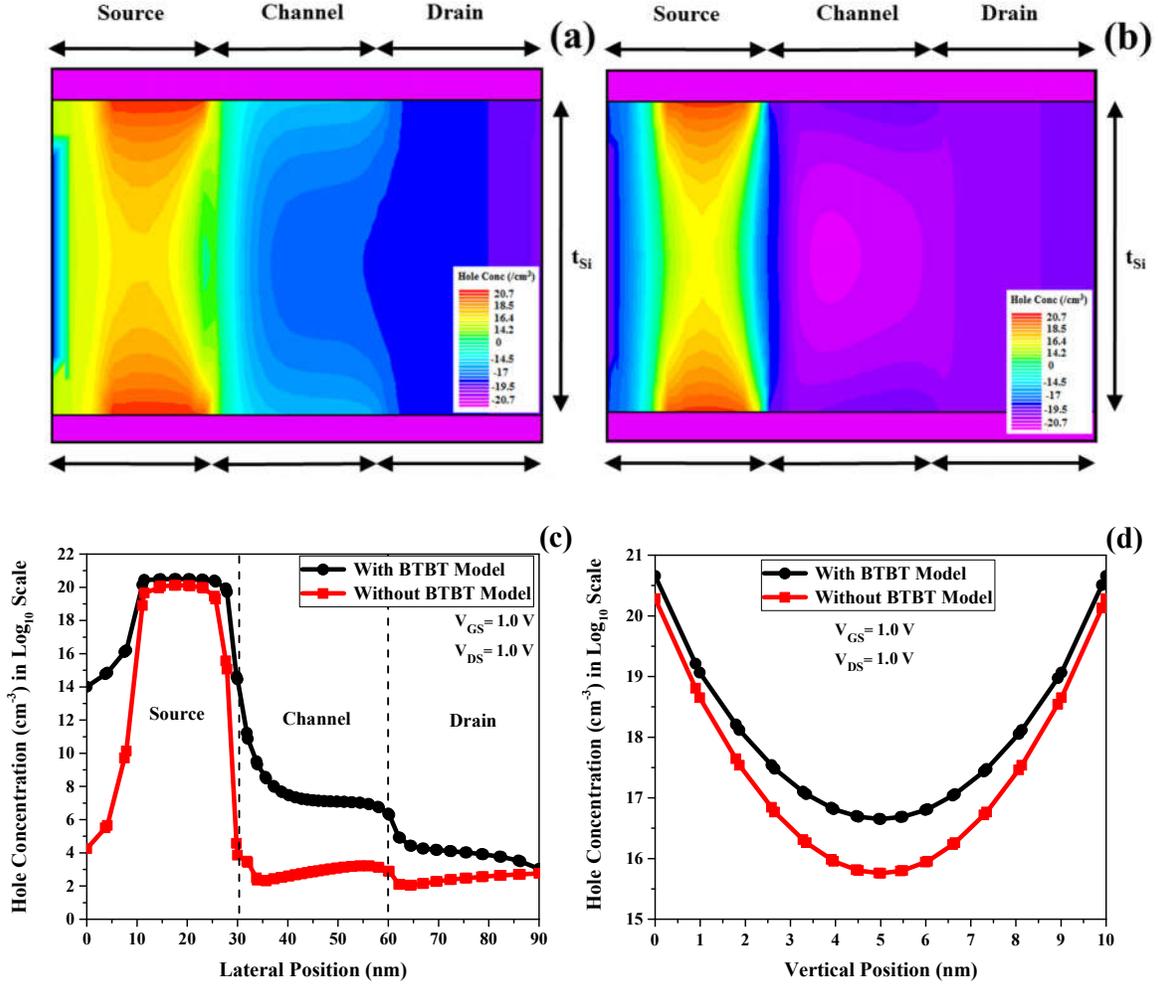

**Fig. 7.** Two-dimensional contour plot for hole concentration of the BJN-TFET structure when the BTBT model is (a) included and (b) not included. Hole concentration of the BJN-TFET structure plotted with and without the BTBT model at $V_{GS}$ = 1.0 V and $V_{DS}$ = 1.0 V in the (c) lateral and (d) vertical directions. There is an increased hole accumulation under the PS-Gate when the BTBT model is included in the simulation.

As can be seen from **Fig. 8**, the excess accumulated hole indeed enhances the potential of the region under the PS-Gate. This leads to a reduction of the barrier between the $n^+/p^+$ regions in the source region, forward biases the base emitter junction, and thereby switches the BJT to the on-state. So the BJT current is merged with the tunneling current and thus provides high drive current of the BJN-TFET structure. Also, **Fig. 8 (c)** and **(d)** demonstrate the potential distribution in the lateral and vertical directions at $V_{GS}$= 1.0 V and $V_{DS}$= 1.0 V which exhibit the increment of the potential under the PS-Gate thanks to the BTBT phenomenon.

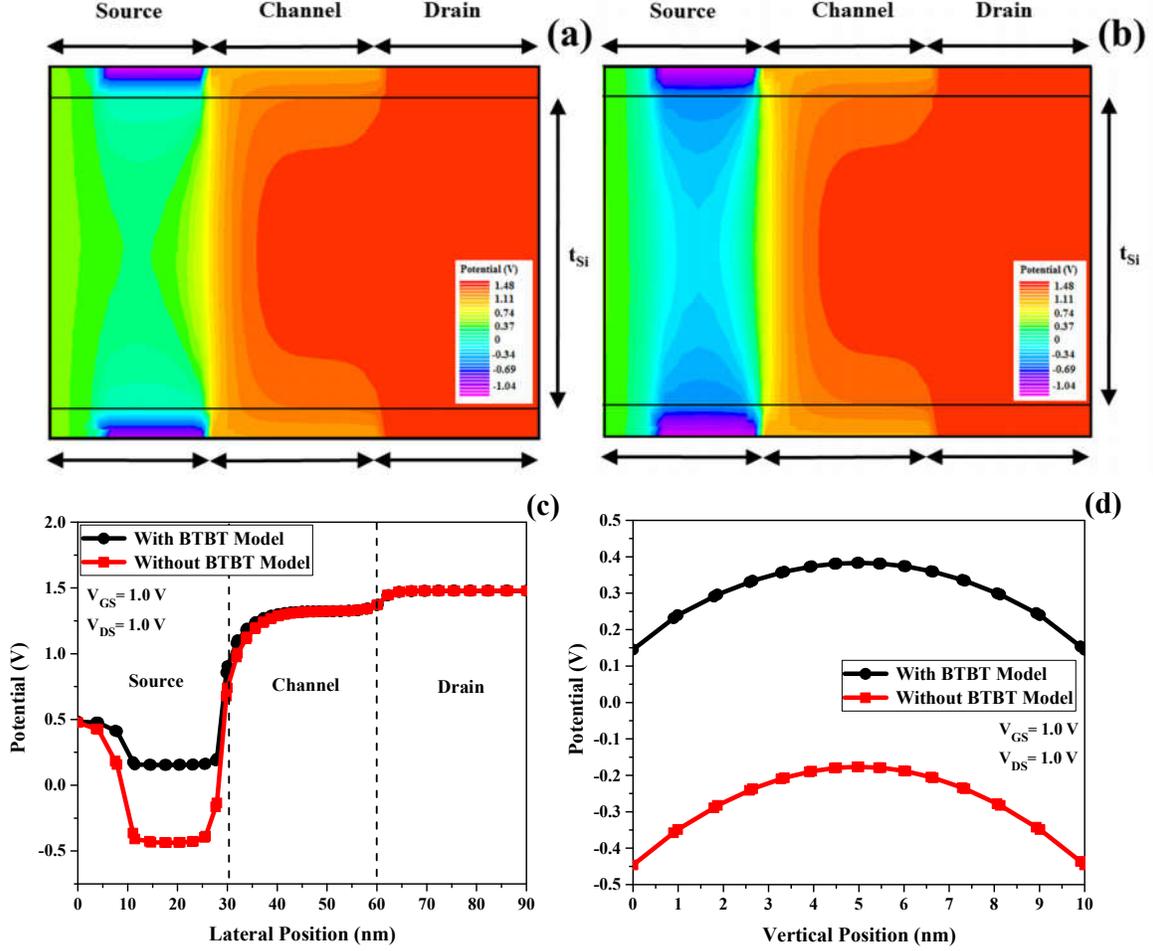

**Fig. 8.** Contour plots of potential distribution in the BJN-TFET structure (a) with and (b) without inclusion of the BTBT model. Two dimensional plot for potential of the BJN-TFET structure when the BTBT model is included and not included at $V_{GS}$ = 1.0 V and $V_{DS}$ = 1.0 V in the (c) lateral and (d) vertical directions. The excess accumulated holes thanks to the BTBT increases the potential under the PS-Gate and thus triggers the BJT device.

### 3.2) Optimization for the BJN-TFET structure

Since in general TFETs suffer from low $I_{on}$, it is desirable to adjust the parameters of the BJN-TFET structure to obtain a maximum $I_{on}$ and also a favorable $I_{on}/I_{off}$. It is always advisable to have a high workfunction for the PS-Gate to move the conduction and valence band edges toward higher energy levels, so that in the on-state a considerable band overlap is achieved between the source and the channel regions for tunneling phenomenon **[43-46]**. **Fig. 9** displays the lateral band diagram of the BJN-TFET structure in the off-state and the corresponding transfer characteristics ($I_{DS}$-$V_{GS}$) for three different PS-Gate workfunction. Lower PS-Gate workfunction leads to off-state current enhancement due to the partial depletion of the region under the PS-Gate and less barrier height between the $n^+$-$p^+$ in the source region. Also, decreased on-state current is because of the band overlap reduction between the source and the channel regions. So, a low PS-Gate workfunction leads to degradation of the on and off state current.

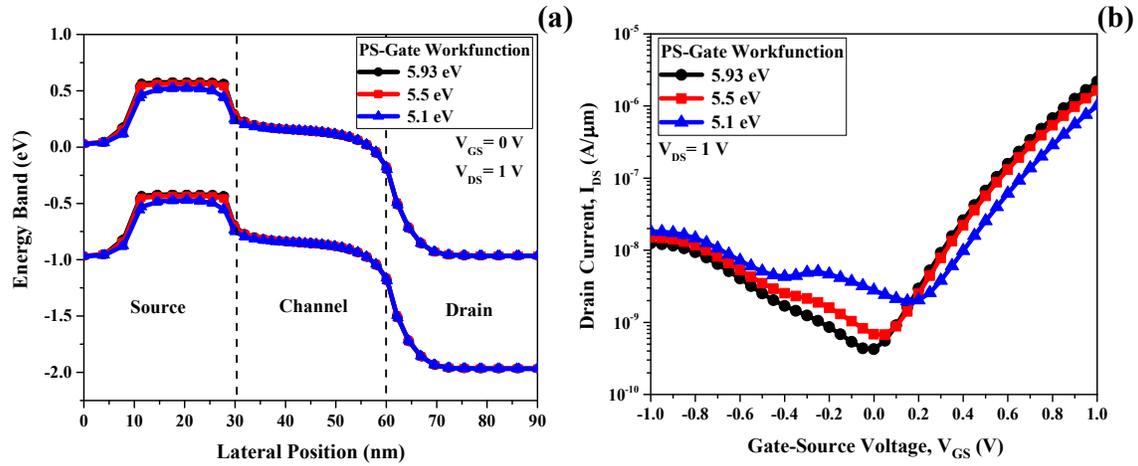

**Fig. 9.** (a) Lateral band diagram of the BJN-TFET structure in the off-state ($V_{GS}$= 0 V and $V_{DS}$= 1 V) for three different PS-Gate workfunction of 5.1, 5.5 and 5.93 eV; (b) $I_{DS}$-$V_{GS}$ characteristics of the BJN-TFET structure at $V_{DS}$= 1.0 V. Off-state current enhancement and on-state current reduction are achieved with lower PS-Gate workfunction.

As shown in **Fig. 10**, the space between the PS-Gate and the main gate has a noticeable impact on the on-state current while it has a negligible effect on the off-state current of the BJN-TFET. The on-state current enhances by approximately one order of magnitude for a 2 nm change in the mentioned space. As we have known, the positive bias on the main gate and high workfunction of the PS-Gate act in opposite directions in the channel and source regions, respectively. In the on-state, the positive bias on the main gate reduces the conduction and valence band energy levels in the channel while owing to the high workfunction of the PS-Gate, the band energy levels are maintained at higher energy levels in the source. Notably, by decreasing the space between the channel and PS-Gate, the barrier height at $n^+/p^+$ interface in the source region as well as the tunneling width between the source and channel regions are reduced. This results in a stronger activation of the built-in BJT device and hence a higher on-state current.

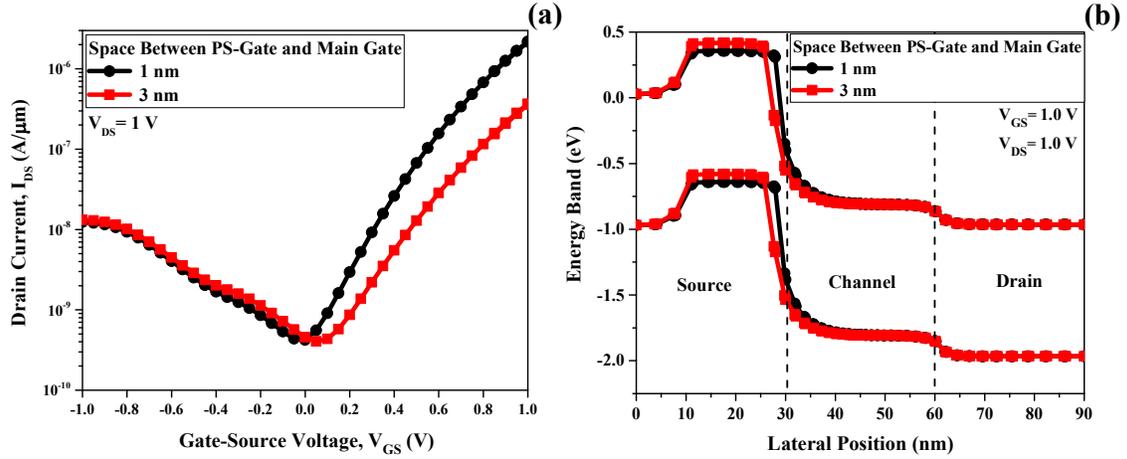

**Fig. 10.** (a) $I_{DS}$-$V_{GS}$ characteristics of the BJN-TFET structure at $V_{DS}$= 1.0 V. The on-state current increases by about one order of magnitude for a 2 nm change in the space between the PS-Gate and the main gate. (b) Lateral band diagram of the BJN-TFET structure in the on-state ($V_{GS}$= 1 V and $V_{DS}$= 1 V) for two space of 1 nm and 3 nm. The barrier height at $n^+/p^+$ interface and the tunneling width are reduced owing to the narrowing of space.

*3.3) Sub-30 nm BJN-TFET*

As we have known, scaling the devices into the sub-20 nm regimes has become a considerable challenge **[39-49, 48, 59-60]**. In TFETs, since the channel current is controlled by the tunneling mechanism in the source side, they are more immune to short-channel effects (SCEs) unlike the conventional nanoscale MOSFETs **[59-61]**. Also, the JN-TFET and dopingless TFET were proposed to serve as attractive approaches for short channel TFETs **[45, 61]** because of their starting junctionless structure which are immune to SCEs **[11-13]**.

As can be seen from **Fig. 11**, two-dimensional device simulations are performed to elucidate the operation of short-channel BJN-TFET and subsequently to examine their applications into the sub-20 nm regimes. The long-channel 30 nm BJN-TFET structure exhibits reasonable switching characteristics with an abrupt on-off switching behavior. Also, the 20 nm BJN-TFET primarily retains the 30 nm current-voltage characteristics and does not considerably degrade from the SCEs. Although in sub-20 nm device the SCEs lead to enhancement of the $I_{off}$ and $SS$, the degradation of electrical characteristics is negligible.

Also **Fig. 11 (b)** and **(c)** demonstrate the dependence of $I_{off}$, $I_{on}/I_{off}$, $SS$, and $V_{TH}$ on the channel length variation. As can be seen from the figure, nearly flat diagram of these parameters shows a slight change of them against channel length variation which exhibits the negligible effect of the SCEs on the BJN-TFET characteristics. Thus the BJN-TFET structure can be scaled down into the sub-20 nm regimes without the influence of the SCEs.

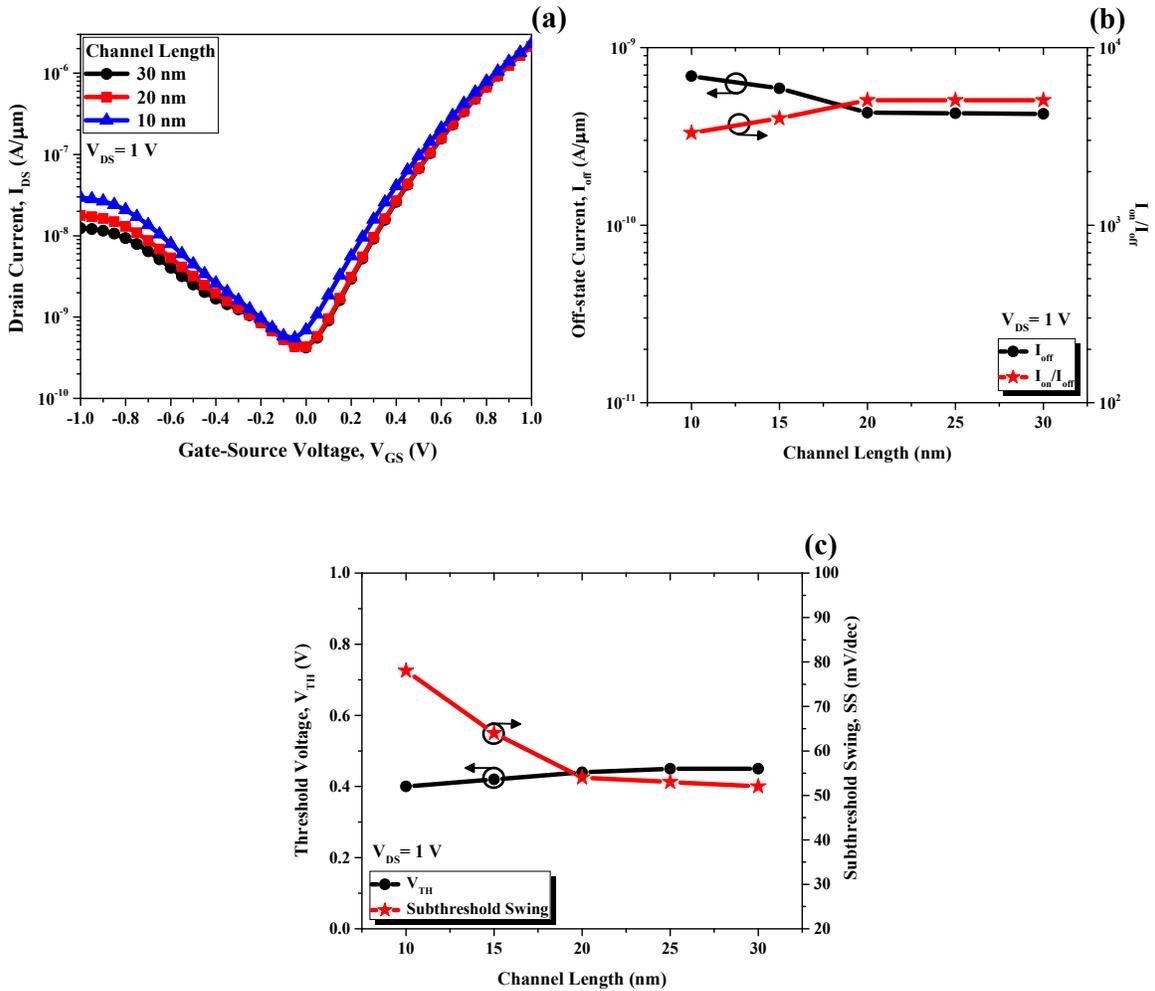

**Fig. 11.** (a) $I_{DS}$-$V_{GS}$ characteristics of the BJN-TFET structure at $V_{DS}$= 1.0 V for various channel length. Enhancement of the off-state current and *SS* values are achieved due to lower channel length. Dependence of (b) $I_{on}$ and $I_{on}/I_{off}$, (c) $V_{TH}$ and subthreshold swing on channel length variation. Low dependence of electrical characteristics on channel length exhibits immunity of the BJN-TFET structure to the SCEs.

## 4. Comparison between the BJN-TFET with TFETs in Literature

**Table. I** exhibits the characteristics of some of the recently published TFETs with improved results along with the results reported in this paper. To further improve the $I_{on}$ and *SS* of a TFET, some innovative techniques are introduced including replacing silicon with lower bandgap semiconductors like Germanium (Ge) and SiGe in the channel **[39-40]**, source **[26-27]**, and a thin layer between the source and channel **[54]**, employing various III–V compounds like InAs and InGaAs **[46]**, strained silicon film **[55-56]**, source pocket with heavily doped **[28-29, 33-34, 52-53]**, dual material gate **[34, 57-58]**, high-k gate dielectric **[30, 32]** and hetero-gate-dielectric **[34-36]**. Worth noting that some of these approaches can be utilized at the cost of significant process complexity **[15, 28-29, 33-37, 52-54]**. According to the table, compared to the other improved devices, the proposed BJN-TFET structure with application of BJT action, without remarkable fabrication challenge thanks to the junctionless starting structure, provides an efficient method to enhance the $I_{on}$ and improve the *SS*. Minimum and maximum values for $I_{on}$ in the below table are $1.5 \times 10^{-8}$ A/µm and $1.0 \times 10^{-3}$ A/µm, respectively, while for the BJN-TFET is $2.1 \times 10^{-6}$ A/µm. Also *SS* of 50 mV/dec in the proposed structure is comparable to the lowest and highest values in the below table which are 22 mV/dec and 225 mV/dec, respectively.

TABLE I

Comparison of Various TFET Devices reported in Literatures and the BJN-TFET Structure in This Paper

| References | Experimental /Simulation | Channel length (nm) | Type | $V_{DS}$ (V) | $V_{GS}$ (V) | SS[1] mV/dec | $I_{on}$[2] A/μm | Dielectric Material | $t_{ox}$ (nm) | Channel material |
|---|---|---|---|---|---|---|---|---|---|---|
| [15] | Experimental | 100 | p | -1.1 | -1.0 | 46 | $1.4 \times 10^{-6}$ | $Al_2O_3$ | 3.5 | Si |
| [16] | Experimental | 70 | n | 1.0 | 1.0 | 52.8 | $1.0 \times 10^{-5}$ | $SiO_2$ | 2 | Si |
| [17] | Simulation | 20 | p | -0.2 | -0.4 | 33.5 | $8.5 \times 10^{-4}$ | $SiO_2$ | 1 | Si |
| [18] | Experimental | 100 | p | -1.1 | -1.0 | 88 | $1.2 \times 10^{-6}$ | $Al_2O_3$ | 3.5 | Si |
| [27] | Simulation | 50 | n | 1.2 | 1.2 | 29 | $1.0 \times 10^{-3}$ | $SiO_2$ | 3 | SiGe/Si |
| [28] | Experimental | 1000 | n | 2.1 | 3.0 | 225 | $4.0 \times 10^{-7}$ | $SiO_2$ | 4 | Bulk Si |
| [28] | Experimental | 1000 | n | 2.1 | 3.0 | 220 | $7.0 \times 10^{-8}$ | $SiO_2$ | 4 | SOI |
| [30] | Simulation | 50 | n | 1.0 | 1.0 | 33 | $8.0 \times 10^{-7}$ | $SiO_2$ | 3 | Si |
| [30] | Simulation | 50 | n | 1.0 | 1.0 | 22 | $5.0 \times 10^{-5}$ | $HfO_2$ | 3 | Si |
| [33] | Simulation | 50 | n | 1.0 | 1.0 | 25 | $1.0 \times 10^{-4}$ | $SiO_2$ | 3 | Si |
| [38] | Simulation | 50 | n | 1.2 | 1.2 | 33 | $1.8 \times 10^{-3}$ | $SiO_2$ | 5 | Ge |
| [39] | Simulation | 50 | p | -0.7 | -0.7 | 28 | $1.3 \times 10^{-6}$ | $HfO_2$ | 3 | SiGe |
| [45] | Simulation | 20 | n | 1.0 | 1.0 | 97 | $1.5 \times 10^{-8}$ | $SiO_2$ | 2 | Si |
| [45] | Simulation | 20 | n | 1.0 | 1.0 | 45 | $8.0 \times 10^{-6}$ | $HfO_2$ | 2 | Si |
| [52] | Simulation | 100 | n | 1.0 | 1.0 | 38 | $6.0 \times 10^{-4}$ | $SiO_2$ | 2.5 | Si |
| [53] | Simulation | 60 | n | 1.0 | 1.0 | 43 | $1.0 \times 10^{-4}$ | $SiO_2$ | 2 | Si |
| [56] | Simulation | 50 | n | 1.0 | 1.0 | 51 | $8.0 \times 10^{-5}$ | $SiO_2$ | 3 | Strained-Si |
| This work | Simulation | 30 | n | 1.0 | 1.0 | 50 | $2.1 \times 10^{-6}$ | $SiO_2$ | 1 | Si |

[1] Steepest *SS*, the best observed subthreshold swing in the $I_{DS}$-$V_{GS}$ characteristics.
[2] $I_{on}$ is measured at the given $V_{DS}$ and $V_{GS}$.

## 5. Conclusion

In summary, we have studied the physical operation of a JN-TFET in detail that combines the merits of the TFET's sharp switching with the high on-current amplification of an n-p-n BJT (BJN-TFET). Some important conclusions which stem from this study regarding the BJN-TFET are as follows: **1.** The BJN-TFET structure is not solely based on tunneling, but also a triggered BJT device has an important role in the on-state; **2.** In the BJN-TFET structure, sub 60 mV/dec subthreshold swing (*SS*) is achieved for low drain currents; **3.** An additional electrode called as PS-Gate in the BJN-TFET structure must have a workfunction as high as possible since it has a strong effect on the on- and off-state current; **4.** For achieving high on-state current, the space between the PS-Gate and the main gate should be thin. So, the design parameters of the BJN-TFET structure are determined in all aspects and this paper acts as a guideline for optimum design of a novel JN-TFET structure which is an attractive alternative for future low power circuit applications.

## Acknowledgment

The authors would like to acknowledge partially support by Nanoelectronic Center of Excellence at department of electrical and computer engineering at University of Tehran.